\documentclass[aps,physrev,twocolumn,groupedaddress,10pt]{revtex4-2}

\usepackage{amsmath}
\usepackage{braket}
\usepackage{bm}
\usepackage{color}
\usepackage{graphicx}
\usepackage[colorlinks,citecolor=blue,linkcolor=red,urlcolor=blue,hyperindex]{hyperref}
\usepackage{url}
\usepackage{booktabs}
\usepackage{diagbox}
\usepackage{multirow}
\usepackage{makecell}

\begin{document}


\title{Resource-Efficient Teleportation of High-Dimensional Quantum Coherence via Initial Phase Engineering}

\author{Long Huang$^{1}$}
\author{Cai-Hong Liao$^{1}$}
\author{Yan-Ling Li$^{1,2}$}
\altaffiliation{liyanling@jxust.edu.cn}
\author{Xing Xiao$^{3}$}
\altaffiliation{xiaoxing@gnnu.edu.cn}
\affiliation{$^{1}$School of Information Engineering, Jiangxi University of Science and Technology, Ganzhou 341000 China\\
$^{2}$Jiangxi Provincial Key Laboratory of  Multidimensional Intelligent Perception and Control, Jiangxi University of Science and Technology, Ganzhou 341000, China\\
$^{3}$School of Physics and Electronic Information, Gannan Normal University, Ganzhou 341000, China}

\date{\today}

\begin{abstract}
High-dimensional quantum systems leverage an expanded Hilbert space to enhance resilience against decoherence and noise. However, standard quantum teleportation is fundamentally limited by the quadratic growth of measurement complexity and high classical communication overhead, requiring the resolution of $d^2$ Bell states and $2\log_2 d$ classical bits. In this study, we propose a resource-efficient high-dimensional coherence teleportation (REHDCT) protocol. By designing $d$ sets of specialized positive operator-valued measure (POVM) bases, our protocol achieves a 50\% reduction in classical communication by utilizing one of the $d$ designed POVM sets, which effectively scales the measurement complexity from $O(d^2)$ to $O(d)$. Furthermore, we demonstrate that by utilizing initial phase engineering to align the target qudit with the measurement basis, theoretically perfect teleportation of quantum coherence can be achieved for arbitrary qudit states. A quantitative robustness analysis reveals that the protocol remains highly resilient to operational errors, maintaining an efficiency above 99.6\% even under a 0.1 rad phase deviation for $d=16$. Our analysis under various noise models (amplitude damping, phase flip, depolarizing, and dit-flip) confirms that high-dimensional systems exhibit an expanding quantum advantage window as dimensionality increases. Notably, under dit-flip noise, perfect coherence teleportation can be restored through the optimal selection of the POVM basis. These findings establish REHDCT as a practical, hardware-friendly framework for resource-efficient quantum communication in future high-dimensional networks.
\end{abstract}


\maketitle

\section{Introduction}
Quantum teleportation \cite{proposeQT}, a groundbreaking concept since the late 20th century, has become a cornerstone of quantum networks by enabling the reconstruction of unknown states at remote locations without direct particle transmission \cite{network1,network2,network3}. Following decades of development, this protocol has been realized across diverse platforms, spanning from integrated chips \cite{chip1,chip2,chip3,chip4} and meter-scale setups \cite{me1,me2,me3} to 100-km optical fiber \cite{OFC} and 1,400-km satellite-to-ground channels \cite{SC}. As the demand for practical applications and complex quantum tasks grows \cite{task1,task2,task3,task5}, high-dimensional quantum systems (qudits) have emerged as an increasingly important frontier. Compared to augmenting the number of entangled particles \cite{mp1,mp2,mp3}, enhancing the dimensionality of individual particles \cite{qudit1,qudit2,qudit3} offers a richer state space with enhanced noise resilience \cite{task4}, more straightforward coherent control \cite{quditEC}, and reduced circuit complexity \cite{comp1,comp2,comp3}. However, a significant resource bottleneck emerges: the standard teleportation of a $d$-dimensional state imposes a quadratic growth in measurement complexity, requiring the resolution of $d^2$ high-dimensional Bell states (HDBSs) and the transmission of $2 \log_2 d$ classical bits. This dimensional scaling leads to a significant increase in resource requirements, posing challenges for hardware scalability, especially in resource-constrained environments. While diverse variants like gate \cite{gateT1,gateT2,gateT3} and port-based teleportation \cite{portT1,portT2,portT3} have expanded the protocol's applicability \cite{QET,QFIT}, there remains a critical need for resource-efficient frameworks specifically tailored to mitigate the escalating physical and informational costs in high-dimensional quantum networking.

Quantum coherence (QC), stemming from the principle of superposition, is the fundamental resource that distinguishes quantum tasks from their classical counterparts \cite{quantify1,quantify2,quantify3}. From a task-oriented perspective, many emerging quantum applications, such as distributed quantum computing \cite{PhysRevA.59.4249} and quantum interferometry \cite{Giovannetti2011}, primarily rely on the successful transmission of coherence rather than the exhaustive reconstruction of an entire quantum state. This motivates a critical inquiry: can the complex objective of full-state teleportation be simplified to a resource-efficient coherence teleportation? While previous studies have demonstrated that a qubit's coherence can be teleported using only one classical bit through specialized positive operator-valued measure (POVM) measurements and initial phase engineering \cite{teleportationQC,noise1}, a natural question arises as to whether analogous resource savings can be achieved for qudit systems, which possess even richer coherence resources within their expanded Hilbert spaces.

In this work, we introduce a general theoretical framework for resource-efficient high-dimensional coherence teleportation (REHDCT). The primary innovation of our protocol lies in the utilization of specialized sets of POVM bases, which allow for a significant reduction in physical and informational costs. Specifically, by substituting standard high-dimensional Bell-state measurements (HDBSMs) with our POVM-based approach, we effectively scale down the measurement complexity from $O(d^2)$ to $O(d)$ possible outcomes and halve the classical communication requirement to only $\log_2 d$ bits. Furthermore, we demonstrate that by incorporating initial phase engineering, the REHDCT protocol can reach the theoretical upper bound of performance, achieving perfect teleportation of quantum coherence for arbitrary qudit states, including both pure and mixed states. Crucially, numerical robustness analysis reveals that the protocol exhibits remarkable resilience against phase engineering deviations, maintaining an average teleportation efficiency exceeding 99.6\% even under typical experimental imperfections. This strategy is particularly hardware-friendly for physical platforms such as orbital angular momentum (OAM) encoding, where distinguishing $d^2$ independent states is often experimentally prohibitive \cite{Erhard2018,PhysRevLett.124.190502}. By grouping detection regions into $d$ outcomes, our scheme lowers the resolution requirements for detection systems, making high-dimensional coherence-based tasks feasible in resource-constrained quantum networks.

In practical quantum teleportation scenarios, qudits inevitably interact with their environment, which can degrade the coherence of the teleported state. Utilizing the completely positive mapping framework of the Choi-Jamiokowski-Kraus-Sudarshan (CJKS) theory \cite{PRA.87.022310}, we model the REHDCT process as a composition of two linear mappings: the environmental noise channel $\mathcal{G}$ and the measurement-induced mapping $\mathcal{J}_E^*$. This unified framework allows us to rigorously evaluate the coherence teleportation efficiency for arbitrary qudit states under several representative noise models: amplitude damping (AD), phase flip (PF), depolarizing (DP), and dit-flip (DF) noise. We derive the noise thresholds at which the quantum advantage vanishes as a function of dimensionality $d$, revealing that the ``quantum advantage window" significantly expands in higher dimensions. Our results confirm that increasing $d$ shifts these thresholds toward higher noise regimes, thereby underscoring the inherent advantages of high-dimensional systems in noise-tolerant tasks.
Furthermore, we observe that under DF noise, perfect coherence teleportation can be achieved by appropriately designing an appropriate POVM basis. These findings on noise resilience, combined with the reduced measurement complexity and classical resource requirements of our protocol, provide a robust solution for implementing high-capacity quantum communication in realistic, noisy environments.

The remainder of this paper is organized as follows: In Sec.~\ref{sec2A}, we establish the theoretical framework of the REHDCT protocol for arbitrary qudit states, demonstrating the reduction of measurement complexity from $O(d^2)$ to $O(d)$. In Sec.~\ref{sec2B}, we show how to achieve the perfect teleportation of QC through initial phase engineering.
In Sec.~\ref{sec2C}, we present a quantitative robustness analysis to evaluate the impact of phase-engineering deviations on teleportation efficiency. Secs.~\ref{sec3A} and \ref{sec3B} are devoted to the derivation of the classical bound for coherence teleportation and a comprehensive performance analysis of our REHDCT protocol under various noise channels, including AD, PF, DP, and DF noise. In Sec.~\ref{sec3C}, we discuss the specific POVM design for restoring perfect coherence teleportation under DF noise. Finally, we conclude the paper in Sec.~\ref{sec4}.

\section{Resource-Efficient Coherence Teleportation in Noiseless High-Dimensional Systems}
\label{sec2}

\subsection{The REHDCT protocol}
\label{sec2A}
In this section, we establish the theoretical framework for the REHDCT protocol. This protocol involves two primary parties: the sender Alice and the receiver Bob. To facilitate the teleportation of QC, they must initially share a maximally entangled state in a $d$-dimensional Hilbert space $\mathcal{H}_d$. The shared maximally entangled state between Alice and Bob can be expressed as
\begin{equation}
\label{eq1}
\ket{\Phi}_{AB}=\sum_{k=0}^{d-1} \frac{1}{\sqrt{d}} \ket{kk},
\end{equation}
where particles $A$ and $B$ are held by Alice and Bob, respectively.

Before the commencement of the teleportation protocol, Alice prepares a target qudit $T$ in an unknown state, which encodes the coherence information she intends to transmit. For an arbitrary qudit state (encompassing both pure and mixed states), the density matrix is expressed as
\begin{equation}
\label{eq2}
    \rho_T=\sum^{d-1}_{j,j'=0} \rho_{jj'} \ket{j} \bra{j'},
\end{equation}
where $\rho_{jj'} = |\rho_{jj'}| e^{i(\phi_j - \phi_{j'})}$ contains the phase factors $\phi_j$. This density matrix satisfies the standard normalization condition ${\rm Tr}(\rho_T)=1$.

To quantify the quantum coherence teleported, we adopt the $l_1$-norm of coherence \cite{quantify1}
\begin{align}
\label{eq3}
    C_{l_1}(\rho)=\sum_{u \neq v}^{d-1} |\rho_{uv}|,
\end{align}
where $\rho$ is the density matrix of the qudit. The $l_1$-norm measure of QC is basis-dependent and adheres to specific constraints that define a valid coherence measure within the framework of quantum resource theory. These constraints are as follows:

(i) Zero coherence for incoherent states: The coherence measure of an incoherent state is strictly zero. This implies that states represented purely in the computational basis without superposition exhibit no quantum coherence.

(ii) Monotonicity under incoherent operations: The coherence measure does not increase under incoherent operations. Incoherent operations, defined as those that map incoherent states to incoherent states, ensure that the coherence of a quantum state cannot be enhanced through operations that do not involve superposition.

(iii) Non-increase on average under selective incoherent operations: The coherence measure must not increase on average when selective incoherent operations are applied. This requirement ensures that the average coherence across all possible outcomes of a quantum operation remains non-increasing, preserving the consistency of coherence as a quantifiable resource.

The central concept of our REHDCT protocol involves substituting the conventional $d^2$ HDBSMs with a single set of specialized POVM measurements comprising $d$ operators. These POVM operators are constructed by partitioning the $d^2$ standard HDBS projectors into $d$ composite elements. The HDBSs are defined as:
\begin{align}
    \label{eq4}
    \ket{\Psi_{nm}}=\sum^{d-1}_{j=0} e^{i2\pi \frac{jn}{d}} \ket{j}\otimes\ket{j\oplus m} /\sqrt{d},
\end{align}
where $\oplus$ denotes addition modulo $d$, i.e., $j\oplus m=(j+m) \text{ mod } d$. We can design $d$ sets of POVM, where the $y$-th element of  the $x$-th POVM set \{$\bm{\Pi}_x$\}=\{$\Pi_x^0, \Pi_x^1, ... , \Pi_x^{d-1}$\} can be written as
\begin{align}
\label{eq5}
    \Pi_x^y=\sum^{d-1}_{l=0} \ket{\Psi_{xl\oplus y,l}} \bra{\Psi_{xl\oplus y,l}}.
\end{align}
By grouping $d$ Bell projectors into a single POVM element, we effectively reduce the number of possible measurement outcomes to $d$. It is straightforward to verify the completeness relation $\sum_{y=0}^{d-1} (\Pi_x^y)^{\dagger} \Pi_x^y = I_d$ with $I_d$ being the identity operator in the $d$ dimensional Hilbert space $\mathcal{H}_d$. This reduction in measurement complexity significantly eases the hardware requirements for high-dimensional platforms, such as OAM encoding, by lowering the resolution threshold of the detection system.

After Alice performs the measurement with $\Pi_x^y$, Bob's particle qudit $B$ collapses to
\begin{align}
    \label{eq6}
    \rho_B^{\Pi_x^y} = \frac{1}{p^{\Pi_x^y}}{\rm Tr} _{AT}\Big[\Pi_x^y\Big(\rho_{T}\otimes\rho_{AB}\Big)(\Pi_x^y)^{\dagger}\Big],
\end{align}
where $p^{\Pi_x^y}={\rm Tr} \Big[\Pi_x^y\Big(\rho_{T}\otimes\rho_{AB}\Big)(\Pi_x^y)^{\dagger}\Big]$ is the corresponding probability of measurement $\Pi_x^y$ and $\rho_{AB} = \ket{\Phi}_{AB}\bra{\Phi}$.

To quantify the performance of the REHDCT protocol, we define a QC teleportation efficiency $\eta$ as the ratio of the coherence received by Bob to the coherence of the initial state
\begin{align}
\label{eq7}
    \eta = \frac{\mathcal{C}_{l_1}(\rho_B^{\Pi_x^y})}{\mathcal{C}_{l_1}(\rho_T)}.
\end{align}

According to Eq. (\ref{eq3}), we can calculate the coherence that Bob received for any initial state $\rho_T$. Here, we directly present the final state to which Bob collapses after Alice applies the measurement operator $\Pi_x^y$
\begin{align}
\label{eq8}
    \rho_B^{\Pi_x^y} &= \sum_{j=0}^{d-1} \frac{1}{d}\rho_{jj} \\
    &+ \sum_{j_1\neq j'_1}^{d-1} \frac{1}{d} \sum_{l=0}^{d-1} e^{i2\pi \frac{(xl\oplus y)(j'_1-j_1)}{d}} \rho_{j_1j_1'} \ket{j_1\oplus l} \bra{j'_1\oplus l}. \nonumber
\end{align}
where $1/d$ is the probability of measurement operator $\Pi_x^y$. The success probabilities are the same for all measurement operators across different basis sets. The details are provided in Appendix \ref{appA}.

Then, the $l_1$-norm measure of its coherence is
\begin{align}
\label{eq9}
    \mathcal{C}_{l_1}(\rho_B^{\Pi_x^y}) = \sum_{j_1 \neq j'_1}^{d-1} \frac{1}{d} \Big|\sum_{l=0}^{d-1} e^{i2\pi \frac{(xl\oplus y)(j_1'-j_1)}{d}} \rho_{j_1\ominus l,j_1'\ominus l}\Big|,
\end{align}
where $\ominus$ denotes subtraction modulo $d$, i.e., $j_1\ominus l = (j_1-l) \text{ mod } d$. We note that the above equation satisfies the following inequality
\begin{align}
\label{eq10}
    \mathcal{C}_{l_1}(\rho_B^{\Pi_x^y}) &\leq \sum_{j_1 \neq j_1'}^{d-1} \frac{1}{d} \sum_{l=0}^{d-1} |\rho_{j_1\ominus l,j_1'\ominus l}| = \mathcal{C}_{l_1}(\rho_T).
\end{align}
If Alice were given a completely unknown state with random phases, she could still perform the REHDCT protocol. The teleportation would remain resource-efficient (using only $\log_2 d$ bits), but the efficiency $\eta$ might be less than unity due to destructive interference.

While standard quantum teleportation facilitates perfect coherence transfer through full-state reconstruction, it imposes a significant resource burden in high dimensions, requiring $d^2$ HDBSMs and $2\log_2 d$ classical bits. In contrast, our REHDCT protocol circumvents this bottleneck by scaling the measurement complexity down from $O(d^2)$ to $O(d)$ outcomes while halving the classical communication overhead to $\log_2 d$ bits. For a system with $d=16$, this translates to distinguishing only 16 POVM outcomes instead of 256 Bell states. Such a reduction substantially alleviates the stringent demands on coincidence-counting electronics and enhances the technical feasibility of implementations on integrated photonic platforms.

\subsection{Perfect Teleportation of QC via Initial Phase Engineering}
\label{sec2B}
In this subsection, we demonstrate that by applying initial phase engineering to the target qudit state $\rho_T$, the coherence can be completely teleported to the receiver. This process ensures that the non-diagonal elements of the density matrix interfere constructively, thereby saturating the theoretical upper bound of coherence teleportation.

According to Eq. (\ref{eq10}), the upper bound  is saturated if and only if all the phase factors in the summation within the mapping align constructively. This condition is satisfied when:
\begin{equation}
\label{phase}
e^{i \left[\frac{2\pi(xl\oplus y)(j_1'-j_1)}{d}+\phi_{j_1\ominus l,j_1'\ominus l}\right]}={\rm const},
\end{equation}
for all values of $l \in \{0, 1, \dots, d-1\}$. In our REHDCT protocol, while the magnitudes of off-diagonal elements $|\rho_{jj'}|$---which determine the $l_1$-norm coherence---remain arbitrary and unknown to Alice, precise initial phase engineering is essential to prevent destructive interference and achieve perfect QC teleportation. The physical intuition behind this process can be compared to polarization matching in classical optics: just as one must align the polarization of light to maximize transmission through a polarizer, our target state must be ``phase-aligned" with the pre-optimized POVM basis. Since the measurement basis is fixed to achieve resource efficiency, this phase engineering acts as a matching process that ensures all off-diagonal elements of the density matrix interfere constructively, thereby saturating the theoretical upper bound of coherence teleportation without altering the nature of the quantum resource itself.

This specific phase engineering is intrinsically linked to the chosen POVM measurement basis. For example, if she selects the POVM \{$\bm\Pi_x$\}, the phase factor of the quantum state should be $\phi_{jj'}=\phi_j-\phi_{j'}$ and $\phi_j = \frac{xj(d-j)}{d}\pi$, then Eq. (\ref{eq9}) can be expressed as
\begin{align}
\label{eq11}
     \mathcal{C}_{l_1}(\rho_B^{\Pi_x^y}) &= \sum_{j_1 \neq j_1'}^{d-1} \frac{1}{d} \Big|e^{i\frac{x(j_1'-j_1)(j_1+j_1'-d)}{d}\pi} \sum_{l=0}^{d-1}  |\rho_{j_1\ominus l, j_1'\ominus l}|\Big| \nonumber\\
     &= \sum_{j_1 \neq j_1'}^{d-1} |\rho_{j_1,j_1'}| = \mathcal{C}_{l_1}(\rho_T).
\end{align}

Experimentally, Alice can implement this phase engineering starting from a reference state $\rho_R$ that possesses no relative phases, expressed as a real density matrix:
\begin{align}
\label{eq12}
     \rho_R = \sum^{d-1}_{j,j'=0} |\rho_{jj'}| \ket{j} \bra{j'},
\end{align}

The target phase profile is then constructed by applying a sequence of $d$ phase gate operations \cite{PG1,PG2,PG3}:
\begin{align}
\label{eq13}
U_{\phi_j} = I_d + (e^{i\pi \frac{xj(d-j)}{d}}-1) \ket{j}\bra{j},
\end{align}
where $U_{\phi_0}=I_d$ corresponds to an identity operation. The resulting engineered state is:
\begin{align}
\label{eq14}
\rho_P= \left(\prod_{j=0}^{d-1} U_{\phi_j} \right) \rho_R \left(\prod_{j=0}^{d-1} U_{\phi_j}^\dagger \right).
\end{align}
Through this systematic preparation, the qudit satisfies the necessary phase conditions to facilitate perfect coherence teleportation.

While perfect, lossless teleportation requires specific initial phase engineering, the core advantage of our scheme lies in its resource efficiency. By focusing on coherence as the primary resource, our protocol requires only $d$ possible measurement outcomes, corresponding to $\lceil \log_{2}d \rceil$ classical bits, where $\lceil \cdot \rceil$ denotes the ceiling function. This represents a 50\% saving in classical communication and a reduction in measurement complexity from $O(d^2)$ to $O(d)$ compared to standard full-state teleportation.

\subsection{\label{sec2C}Robustness Against Initial Phase Engineering Deviations}
In practical experimental implementations, achieving perfect initial phase engineering may be hindered by hardware limitations, such as the finite resolution of spatial light modulators (SLMs) or systematic calibration errors in phase gates.
Here, we evaluate the REHDCT protocol's performance under non-ideal scenarios where the target qudit exhibits operational phase perturbations.

We model these imperfections by assuming that the prepared state $\rho_{\rm error}$ deviates from the ideally engineered state by a small phase factor $\delta_\phi$. This deviation accounts for residual relative phases in the reference state $\rho_R$ and operational inaccuracies in the phase gate array. The resulting state is expressed as:
\begin{align}
\label{eq15}
    \rho_{\rm error} &= \sum^{d-1}_{j=0} |\rho_{jj}| \ket{j} \bra{j} \\
    &+ \left(\sum^{d-1}_{j_1 < j_1'} |\rho_{j_1j_1'}| e^{i(\phi_{j_1j_1'}+\delta_\phi)} \ket{j_1} \bra{j_1'}+ h.c.\right),\nonumber
\end{align}
where $\phi_{j_1j_1'}$ represents the ideal phase difference and $\delta_\phi$ is the perturbation. Following the REHDCT protocol, the coherence received by Bob is given by:
\begin{align}
\label{eq16}
    \mathcal{C}_{l_1}(\rho_E^{\Pi_x^y}) = \sum_{j_1 \neq j_1'}^{d-1} \frac{1}{d} \left| \sum_{l=0}^{d-1}  |\rho_{j_1\ominus l, j_1'\ominus l}| e^{(i \delta_\phi \operatorname{sgn}[(j_1'\ominus l) - (j_1\ominus l)])} \right|,
\end{align}
where $\operatorname{sgn}[x]$ denotes the sign function: it returns $1$ when $x > 0$ and $-1$ when $x < 0$.

Due to the complex correlation between the phase deviation $\delta_\phi$ and the state's density matrix elements, we employ a Monte Carlo approach to quantify the average impact on teleportation efficiency $\eta$. We sample ensembles of qudit states, including pure states sampled via the Haar measure and mixed states via the Hilbert-Schmidt measure \cite{mixedstate}. The average efficiencies for dimensions $d \in \{3, 4, 8, 16\}$ under operational deviations $\delta_\phi \in \{0.01, 0.05, 0.1\}$ rad are summarized in Table \ref{tab1}. Further details are provided in Appendix \ref{appB}.

\begin{table}[htbp]
    \centering
    \caption{Average coherence teleportation efficiency $\eta$ for pure and mixed qudit states under various phase deviations $\delta_\phi$.}
    \label{tab1}
    \begin{tabular}{c|c|c|c}
        \hline
        \diagbox{$d$}{$\delta_\phi$}& \shortstack{0.01 rad\\(pure/mixed)} & \shortstack{0.05 rad\\(pure/mixed)} & \shortstack{0.1 rad\\(pure/mixed)} \\
    \hline
        3 & 0.99996/0.99996 & 0.99901/0.99901 & 0.99608/0.99602 \\
        \hline
        4 & 0.99996/0.99996 & 0.99902/0.99901 & 0.99612/0.99603 \\
        \hline
        8 & 0.99996/0.99996 & 0.99910/0.99909 & 0.99641/0.99636 \\
        \hline
        16 & 0.99996/0.99996 & 0.99913/0.99913 & 0.99653/0.99652 \\
        \hline
    \end{tabular}
\end{table}

The numerical results yield several critical insights into the experimental viability of the REHDCT protocol. Strikingly, even with a phase deviation of $\delta_\phi = 0.1$ rad (equivalent to a significant operational error of approximately $5.7^\circ$), the teleportation efficiency remains remarkably high, exceeding $99.6\%$ across all tested dimensions. Furthermore, the performance is virtually identical for pure and mixed states, confirming the protocol's universal applicability.

Most importantly, the efficiency does not degrade as the dimensionality $d$ increases, suggesting that the REHDCT protocol is inherently scalable. This high tolerance demonstrates that the requirement for initial phase engineering does not pose a prohibitive barrier to experimental realization. Instead, the protocol is robust against typical laboratory imperfections, ensuring that the benefits of reduced measurement complexity and classical resource savings are fully realizable in practical high-dimensional quantum networks.

\section{\label{sec3}Quantum Advantage and Robustness of REHDCT under Environmental Noise}

The primary motivation for transitioning to high-dimensional quantum systems lies in their expanded Hilbert space, which provides superior information capacity and enhanced resilience against environmental decoherence. In this section, we evaluate the performance of REHDCT protocol by analyzing its performance in various noisy environments. It is important to emphasize that throughout the following noise analysis, we assume that the target state has been prepared using the initial phase engineering described in Eq. (\ref{phase}).

To quantify the quantum advantage, we derive the classical bound on coherence teleportation without entanglement and define the \textit{efficiency of coherence teleportation} ($\eta_{l_1, \rm cl}$) as the ratio between the teleported coherence received by Bob and the initial coherence prepared by Alice. This benchmark allows us to precisely identify the regime where our REHDCT protocol outperforms any classical strategy: a quantum advantage exists if and only if the teleportation efficiency $\eta$ exceeds the classical threshold $\eta_{l_1, \rm cl}$. By comparing the REHDCT efficiency against this benchmark across several representative noise models---including AD, PF, DP, and DF noise---we demonstrate that high-dimensional systems not only preserve quantum advantages under severe noise but also exhibit an expanding ``advantage window" as the dimensionality $d$ increases. This analysis underscores the practical significance of the REHDCT protocol for building scalable and noise-tolerant quantum communication networks.

\subsection{The classical bound on the efficiency of QC teleportation}
\label{sec3A}

The optimal classical strategy for the teleportation of QC involves a covariant measurement combined with state re-preparation. This approach assumes that the unknown pure input state is uniformly distributed on the Bloch sphere, and it entails averaging over both the input distribution and the measurement outcomes. However, this approach is not readily extendable to high-dimensional scenarios. In this context, we adopt an alternative strategy to derive the classical bound of QC teleportation from the perspective of noisy singlets. Noisy singlets are the most natural generalization of the 2-dimensional Werner states \cite{classicT}.
\begin{align}
\label{eq17}
    \rho_r=r\rho_{AB} + (1-r)\frac{I_d\otimes I_d}{d^2},
\end{align}
where $r$ represents the purity parameter. Such a state is the only one invariant under $U\otimes U^{\dagger}$ transformations \cite{distillation}. Moreover, the state $\rho_r$ is separable when $0 \leq r \leq \frac{1}{d+1}$, and entangled when $\frac{1}{d+1} < r \leq 1$.

Note that the second term of Eq. (\ref{eq17}) represents the completely random noise that eliminates the coherence of the initial state, irrespective of the nature of the initial state. After Alice performs the measurement with $\Pi_x^y$, Bob's qudit $B$ collapses to
\begin{align}
    \rho_{B,r}^{\Pi_x^y} &= \frac{1}{p^{\Pi_x^y}}{\rm Tr} _{TA}\Big[\Pi_x^y\Big(\rho_{T}\otimes\rho_r\Big)(\Pi_x^y)^{\dagger}\Big] \nonumber\\
    &= \frac{r}{p^{\Pi_x^y}}{\rm Tr} _{TA}\Big[\Pi_x^y\Big(\rho_{T}\otimes\rho_{AB}\Big)(\Pi_x^y)^{\dagger}\Big] \nonumber\\
    &+ \frac{1-r}{p^{\Pi_x^y}}{\rm Tr} _{TA}\Big\{\Pi_x^y\Big[\rho_{T}\otimes \frac{I_d}{d} \otimes \frac{I_d}{d}\Big)\Big](\Pi_x^y)^{\dagger}\Big\} \nonumber\\
    &= r  \rho_{B}^{\Pi_x^y} + (1-r)  \frac{I_d}{d}.
\end{align}

Since the second term $(1-r)  \frac{I_d}{d}$ does not contribute to the QC, we can conclude that the final teleported coherence is given by
\begin{align}
    \mathcal{C}_{l_1}(\rho_{B,r}^{\Pi_x^y})=r \mathcal{C}_{l_1}(\rho_T).
\end{align}
It is important to note that only the purity parameter $r$ determines the proportion of initial QC that can be teleported. The coherence that can be teleported through a purely classical channel in the absence of entanglement is given by
\begin{align}
    \mathcal{C}_{l_1,\rm cl}\leq\frac{1}{d+1}\mathcal{C}_{l_1}(\rho_T).
\end{align}
Therefore, the classical efficiency of QC teleportation is bounded by
 \begin{align}
    \eta_{l_1,\rm cl}=\frac{1}{d+1}.
\end{align}
It is observed that as the dimensionality increases, the classical bound on the efficiency of QC teleportation gradually decreases, thereby facilitating the manifestation of advantages in high-dimensional quantum systems.

\subsection{Performance of REHDCT under Environmental Noise}
\label{sec3B}

\begin{figure}
    \centering
    \includegraphics[width=1\linewidth]{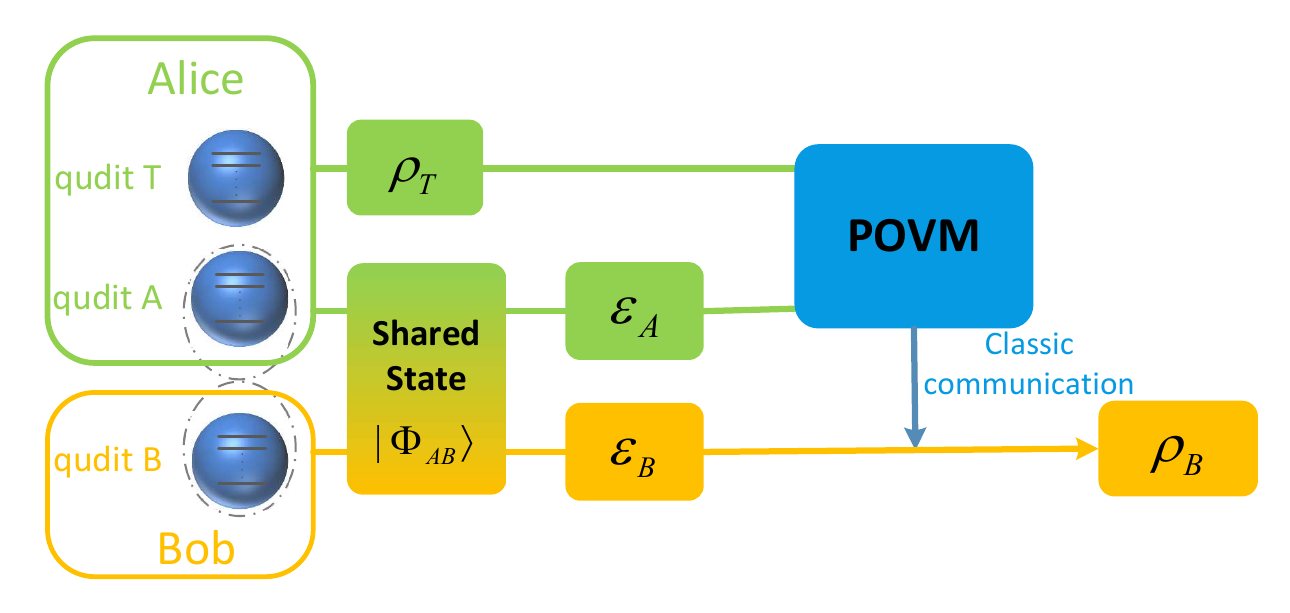}
    \caption{(Color online) Schematic representation of the REHDCT protocol under environmental noise. The target qudit $T$ undergoes initial phase engineering to ensure the phase-alignment condition (\ref{phase}) is satisfied prior to the protocol. The shared maximally entangled state $\rho_{AB}$ is subjected to local noise channels $\mathcal{E}_{A}$ and $\mathcal{E}_{B}$. Alice performs a resource-efficient measurement utilizing the POVM set $\{\Pi_x\}$, which effectively scales the measurement complexity down to $O(d)$.}
    \label{protocol}
\end{figure}

In realistic quantum communication scenarios, systems inevitably interact with their surrounding environment, leading to noise and decoherence that can degrade the quality of resource teleportation. In this section, we evaluate the performance of the REHDCT protocol under such practical constraints. We consider a scenario where particles $A$ and $B$, which initially form a maximally entangled pair, are subjected to independent but identical local noise processes, represented by a quantum channel $\mathcal{E}$. This assumption of identical noise channels allows us to focus on the fundamental impact of environmental decoherence on high-dimensional coherence teleportation without introducing additional complexity from differing noise sources. The protocol for teleporting QC through such noisy channels is illustrated in Fig. \ref{protocol}.

The theoretical framework for teleportation of QC is grounded in the CJKS theorem concerning completely positive maps \cite{CJKS1,CJKS2,CJKS3,CJKS4,teleportationQC}, which provides a robust mechanism for analyzing the effects of noise on the teleportation process. The theory can be succinctly described as follows: Alice possesses an unknown quantum state $\rho_{T}$ and shares an entangled state $\tau$ with Bob. To facilitate the teleportation process, Alice performs a measurement using a POVM element $E$. As a result of this measurement, the state of the particle held by Bob collapses to
\begin{equation}
\label{eq22}
\rho_B=\frac{\mathcal{G}\mathcal{J}_E^*(\rho_{T})}{{\rm Tr} [\mathcal{G}\mathcal{J}_E^*(\rho_{T})]},
\end{equation}
where ${\rm Tr} [\mathcal{G}\mathcal{J}_E^*(\rho_{T})]$ represents the probability associated with the measurement outcome. In this context, $\mathcal{G}$ and $\mathcal{J}_E$ are completely positive maps associated with $\tau$ and $E$, respectively, which are given by
\begin{align}
\label{eq23}
    \tau &=\sum_{i,j=0}^{d-1} e_{ij}\otimes \mathcal{G}(e_{ij}),\\
    E &=\sum_{i,j=0}^{d-1} e_{ij}\otimes \mathcal{J}_E(e_{ij}),
\label{eq24}
\end{align}
where $\{e_{ij} = \ket{i}\bra{j}\}_{i,j=1}^n$ constitutes a complete set of matrix elements for $\mathcal{B}(\mathcal{H})$, with $\mathcal{B}(\mathcal{H})$ denoting the vector space of bounded linear operators on the Hilbert space $\mathcal{H}$.

Note that the central idea of Eq. (\ref{eq22}) is to equivalently represent the noise and measurement processes occurring during quantum teleportation as two distinct mappings, denoted as $\mathcal{G}$ and $\mathcal{J}_E^*$, respectively.
It provides a novel method for calculating the final state obtained by Bob, which is applicable to both the noiseless and noisy scenarios. In the noiseless case ($\tau=\rho_{AB}$),  $\mathcal{G}$ represents $1/d$ times the identity mapping. Conversely, in the presence of noise, the specific form of $\mathcal{G}$ must be determined based on the type of noise, as will be discussed below.

The effect of noisy channel on the entangled state can be described as a Choi map
\begin{align}
\label{eq25}
    \mathcal{E}_{AB}(\rho_{AB}) = (\mathcal{E}_A\otimes \mathcal{E}_B)(\ket\Phi_{AB}\bra{\Phi}).
\end{align}
We now need to convert the bilateral noise mapping into an isomorphic and equivalent one-sided noise mapping to align with the form presented in Eq. (\ref{eq23}). We note that, for the maximally entangled state $\rho_{AB}$ and a completely positive trace-preserving (CPTP) map $\mathcal{E}$, the following property holds (see the details in Appendix \ref{appC})
\begin{align}
\label{eq26}
   (\mathcal{E}_A \otimes \mathcal{I})(\ket{\Phi}_{AB}\bra{\Phi}) = (\mathcal{I} \otimes \mathcal{E}_A^T)(\ket{\Phi}_{AB}\bra{\Phi}),
\end{align}
where $\mathcal{E}^T$ denotes the transpose of the map $\mathcal{E}$ and $\mathcal{I}$ is the identity map. With Eq. (\ref{eq26}) at hand, the Eq. (\ref{eq25}) can be re-written as
\begin{align}
\label{eq27}
    \mathcal{E}_{AB}(\rho_{AB}) &= (\mathcal{E}_A\otimes \mathcal{E}_B)(\ket\Phi_{AB}\bra{\Phi}) \nonumber\\
    &= (\mathcal{I}\otimes \mathcal{E}_B)(\mathcal{E}_A \otimes \mathcal{I})(\ket\Phi_{AB}\bra{\Phi}) \nonumber\\
    &= (\mathcal{I}\otimes \mathcal{E}_B)(\mathcal{I} \otimes \mathcal{E}^T_A)(\ket\Phi_{AB}\bra{\Phi}) \nonumber\\
    &= (\mathcal{I}\otimes (\mathcal{E}_B \circ \mathcal{E}_A^T))(\ket\Phi_{AB}\bra{\Phi}),
\end{align}
where $\circ$ denotes the function composition.

We can derive the noise mapping $\mathcal{G}$ associated with the shared state $\rho_{AB}$ as follows
\begin{equation}
\label{eq28}
\mathcal{G}(\rho) = \frac{1}{d}\mathcal{E}_B \circ \mathcal{E}_A^T(\rho),
\end{equation}
where $\rho$ represents a single-qudit state within the vector space $\mathcal{B}(\mathcal{H})$. Therefore, in the scenario where the shared entangled state is a maximally entangled state, the bilateral noise mapping can be converted into an isomorphic and equivalent one-sided noise mapping.

In the field of open quantum systems, numerous methodologies have been developed to investigate the evolution of system states \cite{noise1,noise2,noise3,noise4,noise5}. In this work, we employ the widely adopted Kraus operator formalism to analyze the effects of noise in the protocol. The Kraus operator formalism describes a trace-preserving map given by $\rho \to \rho'=\sum_k E_k \rho E_k^{\dagger}$, where the Kraus operators $E_k$ taking on different forms corresponding to various noise environments, and satisfies $\sum_k E_k^{\dagger} E_k = I_d$. Consequently, we utilize the Kraus operator representation to express the aforementioned mapping, which can be formulated as
\begin{align}
\label{eq29}
    \mathcal{G}(\rho) &= \frac{1}{d} \mathcal{E}_B \circ \mathcal{E}^T_A(\rho) \nonumber\\
    &= \frac{1}{d} \sum_{a,b}(E_a E_b^T) \rho (E_a E_b^T)^\dagger \nonumber\\
    &= \frac{1}{d} \sum_{a,b}F_{ab} \rho F_{ab}^\dagger,
\end{align}
where $F_{ab}=E_a E_b^T$, $E_a$ and $E_b$ represent the Kraus operators of the noise channels $\mathcal{E}_B$ and $\mathcal{E}_A$, respectively. This formulation facilitates the analysis of the noisy protocol and allows us to introduce an approach for investigating noise.

In our protocol for the teleportation of QC, Alice performs a measurement using a POVM element $\Pi_x^y$. Consequently, it is essential to determine the measurement mapping $\mathcal{J}_E^*$ that corresponds to the operator $\Pi_x^y$. Thus, Eq. (\ref{eq5}) can be written as
\begin{align}
\label{eq30}
    \Pi_x^y &= \sum^{d-1}_{l,j,j'=0} \frac{1}{d}\ket{j}\bra{j'} \otimes e^{i2\pi \frac{j(xl\oplus y)}{d}}\ket{j\oplus l}\braket{j|j} \nonumber\\
    &\hspace{1.5cm} \braket{j'|j'} e^{-i2\pi \frac{j'(xl\oplus y)}{d}}\bra{j'\oplus l},
\end{align}
Then we can get the measurement mapping $\mathcal{J}_E^*$ which is expressed as
\begin{align}
\label{eq31}
\mathcal{J}_E^*(\rho)=\sum_{l=0}^{d-1} (W_{xl\oplus y,l})  \rho  (W_{xl\oplus y,l})^\dagger,
\end{align}
with $W_{nm} = \sum_{j=0}^{d-1} \frac{1}{\sqrt{d}}e^{-i2\pi \frac{jn}{d}}\ket{j\oplus m}\bra{j}$.

By combining Eqs. (\ref{eq22}), (\ref{eq29}) and (\ref{eq31}), we can obtain the results of QC teleportation under different noise environments
\begin{align}
    \rho_{B,X}^{\Pi_x^y} = \frac{\sum_{a,b} F_{ab}^{X} \Big(\sum_{l=0}^{d-1} (W_{xl\oplus y,l})\rho_T(W_{xl\oplus y,l})^\dagger \Big) (F_{ab}^{X})^\dagger}{{d \rm Tr}[\mathcal{G}_{X}\mathcal{J}_E^*(\rho_T)]},
    \label{eq32}
\end{align}
with $X$ denoting noise types and ${\rm Tr}[\mathcal{G}_{X}\mathcal{J}_E^*(\rho_T)]$ the normalization factor corresponding to measurement success probability. In the subsequent discussions, we will examine four representative types of noise, specifically $X \in \{\rm AD, \rm PF, \rm DP, \rm DF\}$.

\subsubsection{AD noise}

AD noise is primarily used to describe energy dissipation phenomena in quantum systems, such as spontaneous emission, photon loss, etc. For simplicity, we assume that the $d-1$ excited states of the qudit decay to the ground state with the same probability.
The Kraus operators can be written as \cite{noise2, noise3}
\begin{align}
    &E_0 = \ket{0}\bra{0} + \sqrt{1-p}\sum_{j=1}^{d-1}\ket{j}\bra{j}, \\
    &E_j = \sqrt{p}\ket{0}\bra{j};
\end{align}
with $j=1, ..., d-1$.

We then calculate the corresponding operator $F_{ab}^{\rm AD}$
\begin{align}
    F_{ab}^{\rm AD} = \left\{
    \begin{aligned}
    &\ket{0}\bra{0} + \sum_{j=1}^{d-1}(1-p) \ket{j}\bra{j} && \text{for } a=0,b=0; \\
    &\sqrt{p(1-p)}\ket{b}\bra{0} && \text{for } a=0,b>0; \\
    &\sqrt{p(1-p)}\ket{0}\bra{a} && \text{for } a>0,b=0; \\
    &p\ket{0}\bra{0} && \text{for } a=b>0; \\
    &0 && \text{for } else.
    \end{aligned}
    \right.
\end{align}

After Alice performs the measurement using the operator $\Pi_x^y$, Bob's particle qudit $B$ collapses to (See Appendix \ref{appD1} for more details)
\begin{widetext}
\begin{align}
    \rho_{B,\rm AD}^{\Pi_x^y} &= \frac{1}{d}\Big[\Big(1+p(d-1)\Big)\ket{0}\bra{0} + \sum_{j=1}^{d-1} (1-p) \ket{j}\bra{j} + \sum_{l=0}^{d-1} \Big( \sum_{j'_1=1}^{d-1} (1-p) e^{i2\pi\frac{j'_1(xl\oplus y)}{d}} \rho_{d\ominus l,j'_1\ominus l} \ket{0}\bra{j'_1} \nonumber\\
        &+ \sum_{j_1=1}^{d-1} (1-p) e^{i2\pi\frac{-j_1(xl\oplus y)}{d}} \rho_{j_1\ominus l,d\ominus l} \ket{j_1}\bra{0} + \sum_{j_2\neq j_2'=1}^{d-1} (1-p)^2 e^{i2\pi\frac{(j'_2-j_2)(xl\oplus y)}{d}} \rho_{j_2\ominus l,j'_2\ominus l} \ket{j_2}\bra{j'_2} \Big)\Big],
\end{align}
\end{widetext}
where $\frac{1}{d}$ is the probability of measurement operator $\Pi_x^y$. The $l_1$-norm coherence is obtained as
\begin{align}
    \mathcal{C}_{l_1}(\rho_{B,\rm AD}^{\Pi_x^y}) = \frac{2(1-p)+(d-2)(1-p)^2}{d} \mathcal{C}_{l_1}(\rho_T).
\end{align}
The efficiency of QC teleportation in this case is given by
\begin{align}
    \eta_{l_1,\rm AD}= \frac{2(1-p)+(d-2)(1-p)^2}{d}.
\end{align}

\begin{figure}[ht]
\centering
\includegraphics[width=\linewidth]{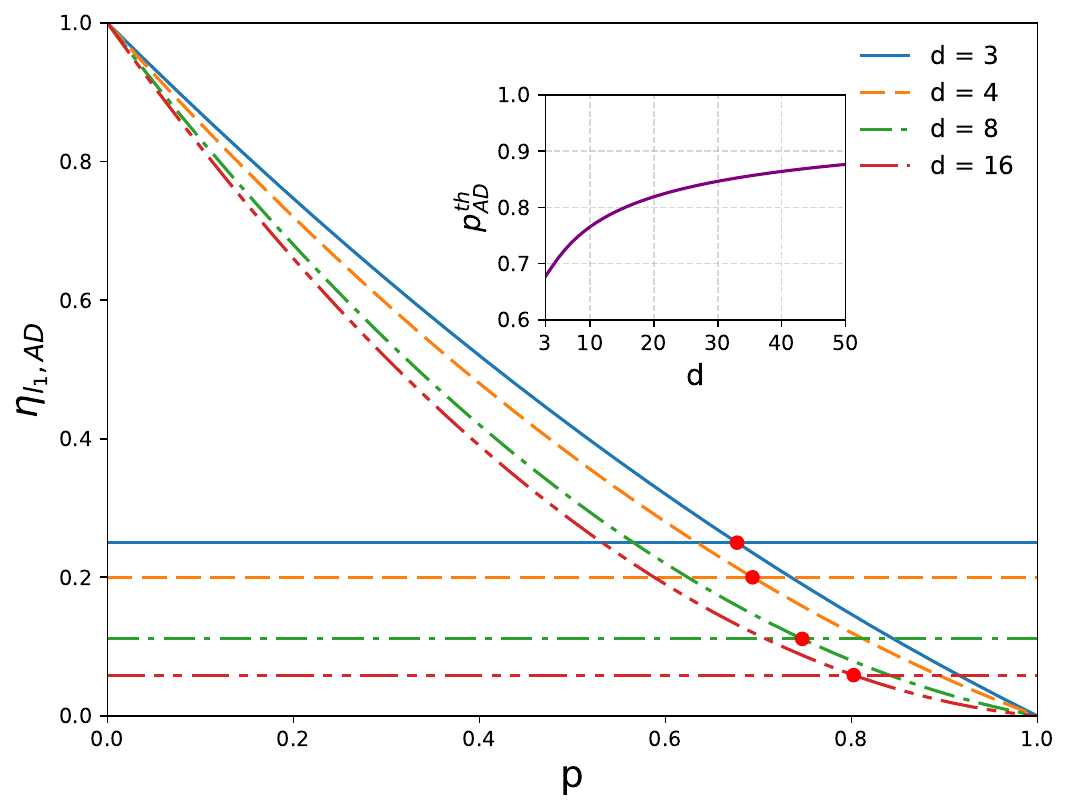}
\caption{(Color online) The efficiency of QC teleportation for qudits of varying dimensions under AD noise. The horizontal lines corresponds to the classical bounds. The red dots indicate the positions of $p_{\rm AD}^{\rm th}$. The inset: The noise threshold $p_{\rm AD}^{\rm th}$ as a function of dimensionality $d$.}
\label{AD}
\end{figure}

Figure \ref{AD} illustrates the teleportation efficiency $\eta$ for qudits of varying dimensions under the influence of AD noise. The observed efficiency trends exhibit behavior analogous to the fidelity decay reported in standard high-dimensional teleportation \cite{noise2}. As the dimensionality $d$ increases, the proliferation of available decay channels from multiple excited states accelerates the overall loss of coherence, indicating that higher-dimensional systems possess inherent sensitivity to AD noise in terms of teleportation efficiency. However, this accelerated decay does not compromise the utility of qudit systems. On the contrary, the quantum advantage is significantly preserved because the corresponding classical benchmark $\eta_{l_1, \rm cl}$---the maximum coherence transferred without entanglement---decreases more rapidly with increasing dimensionality.

To quantify this resilience, we derive the noise threshold $p_{\rm AD}^{\rm th}$ at which the quantum advantage vanishes (i.e., when $\eta$ drops to the classical bound) as a function of the system dimension $d > 2$:
\begin{align}
p_{\rm AD}^{\rm th}= \frac{d^2 - 1 - \sqrt{d^3 + 1}}{d^2 - d - 2}.
\label{eq:placeholder}
\end{align}
Crucially, as illustrated in the inset of Fig. \ref{AD}, the noise threshold $p_{\rm AD}^{\rm th}$ monotonically approaches 1 as $d$ increases. This observation provides a powerful justification for the REHDCT protocol: while high-dimensional states are more fragile in an absolute sense, they maintain a demonstrable quantum advantage under increasingly permissive noise conditions. These results underscore that dimensional scaling effectively broadens the operational window for quantum information tasks, making the REHDCT protocol highly robust for future large-scale quantum networks.

\subsubsection{PF noise}
The PF noise of the qudit can be physically attributed to the random phases induced by the dispersive coupling between the energy levels of the system and the surrounding environment. This phenomenon is effectively equivalent to the application of discrete phase rotations with a specified probability, while preserving the occupation of the energy levels.
Therefore, the random process can be described as follows: With a probability of $(1- p)$, the system remains in the identity operation $I_d$ (indicating no phase change); With a probability of $\frac{p}{d-1}$, a random non-trivial phase rotation $Z^1,\dots,Z^{(d-1)}$ is applied. The corresponding Kraus operators can be expressed as \cite{noise2, noise4}
\begin{align}
    E_{0} = \sqrt{1-p}I_{d},
    E_{m} = \sqrt{\frac{p}{d-1}} Z^m;
\end{align}
where $Z^m = \sum_{j=0}^{d-1}\omega_{d}^{jm}\ket{j}\bra{j}$ and $\omega_d=e^{i\frac{2\pi}{d}}$ with $m=1, ..., d-1$.

We then calculate the corresponding operator $F^{\rm PF}_{ab}$
\begin{align}
    F_{ab}^{\rm PF} = \left\{
    \begin{aligned}
    & a_0^2 \sum_{j=0}^{d-1} \ket{j}\bra{j} && \text{for } a=0,b=0; \\
    & a_0 a_b \sum_{j=0}^{d-1} \omega_d^{bj} \ket{j}\bra{j} && \text{for } a=0,b>0; \\
    & a_0 a_a \sum_{j=0}^{d-1} \omega_d^{aj} \ket{j}\bra{j} && \text{for } a>0,b=0; \\
    & a_a a_b \sum_{j=0}^{d-1} \omega_d^{(a+b)j} \ket{j}\bra{j} && \text{for } a>0,b>0,
    \end{aligned}
    \right.
\end{align}
where $a_0=\sqrt{1-p}$ and $a_m=\sqrt{\frac{p}{d-1}}$.

Subsequently, following Alice's measurement utilizing the operator $\Pi_x^y$, Bob's particle qudit $B$ undergoes a collapse to the state (See Appendix \ref{appD2} for more details)
\begin{align}
    \rho_{B,\rm PF}^{\Pi_x^y} &= \sum_{j=0}^{d-1} \frac{1}{d} \ket{j}\bra{j} + \sum_{j_1\neq j_2=0}^{d-1} \sum_{l=0}^{d-1} \frac{1}{d}(1-\frac{d}{d-1}p)^2 \nonumber\\
    &\hspace{1cm}e^{i2\pi \frac{(j_2-j_1)(xl\oplus y)}{d}} \rho_{j_1\ominus l,j_2\ominus l} \ket{j_1}\bra{j_2},
\end{align}
where $\frac{1}{d}$ is the probability of measurement operator $\Pi_x^y$. The $l_1$-norm coherence yields to
\begin{align}
    \mathcal{C}_{l_1}(\rho_{B,\rm PF}^{\Pi_x^y}) &= (1-\frac{d}{d-1}p)^2 \mathcal{C}_{l_1}(\rho_T).
\end{align}
The resulting efficiency of QC teleportation in PF noise reduces to
\begin{align}
\label{eq44}
    \eta_{l_1,\rm PF}=(1-\frac{d}{d-1}p)^2.
\end{align}

\begin{figure}[ht]
\centering
\includegraphics[width=1\linewidth]{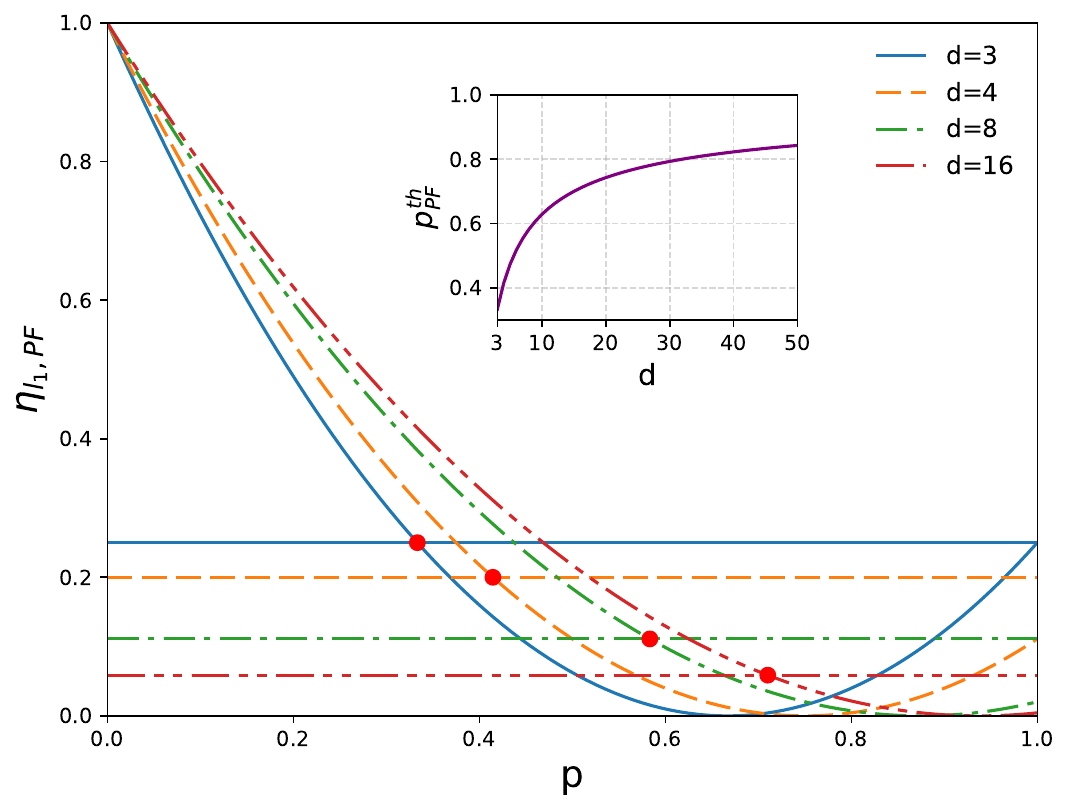}
\caption{(Color online) The efficiency of QC teleportation for qudits of varying dimensions under PF noise. The horizontal lines corresponds to classical bounds. The red dots indicate the positions of $p_{\rm PF}^{\rm th}$. The inset: The noise threshold $p_{\rm PF}^{\rm th}$ as a function of dimensionality $d$.}
\label{PF}
\end{figure}

The performance of the REHDCT protocol under PF noise is summarized in Fig. \ref{PF}. In contrast to the monotonic decay observed under AD noise, the teleportation efficiency $\eta$ exhibits a non-monotonic behavior, characterized by a partial recovery as the noise parameter $p$ approaches 1. This phenomenon arises due to the specific symmetry of the PF channel in qudit systems, where strong noise can effectively act as a coherent phase transformation that partially preserves the resource structure. This non-monotonic recovery as $p \to 1$ can be analytically attributed to the cyclic symmetry of the generalized Pauli $Z$ operators. In the strong noise limit, the off-diagonal elements are scaled by a factor of $-1/(d-1)$ rather than being eliminated, thus partially preserving the coherence resource under the $l_1$-norm measure. However, this recovery effect diminishes significantly as the dimensionality $d$ increases. For instance, at $d=3$, the recovered efficiency in the high-noise regime still fails to surpass the classical benchmark $\eta_{l_1, \rm cl}$. Despite the suppression of this local recovery in higher dimensions, the fundamental advantage of high-dimensional REHDCT remains robustly preserved. Based on the initial phase engineering established in Sec.~\ref{sec2B}, we derive the noise threshold $p_{\rm PF}^{\rm th}$ at which the quantum advantage vanishes:
\begin{align}
\label{eq45}
p_{\rm PF}^{\rm th}=\frac{d-1}{d} \left( 1 - \frac{1}{\sqrt{d + 1}} \right).
\end{align}

As shown in the inset of Fig. \ref{PF}, $p_{\rm PF}^{\rm th}$ monotonically approaches 1 as $d$ increases, mirroring the trend observed in the AD noise analysis. This reinforces our core finding: while high-dimensional systems may exhibit different decay dynamics under various noise models, the range of noise parameters sustaining a quantum advantage increases with dimensionality $d$. This scalability ensures that REHDCT can maintain a superior performance over classical strategies under increasingly relaxed noise constraints.

\subsubsection{DP noise}

Depolarizing noise describes a process in which a qudit is projected into the completely mixed state with a probability of $p$. Its Kraus operators can be written as \cite{noise2, qudit1}
\begin{align}
    &E_{00} = a_{00} \sum_{j=0}^{d-1} \ket{j}\bra{j}, \\
    &E_{mn} = a_{mn} \sum_{j=0}^{d-1} \omega_d^{jm} \ket{j}\bra{j \oplus n};
\end{align}
where $(m,n)\in S$ with $S = \{(a,b) \mid 0 \leq a,b \leq d-1, (a,b) \neq (0,0)\}$, and the coefficients are defined by
\begin{align}
    &a_{00}=\sqrt{1-\frac{d^2-1}{d^2}p}, \\
    &a_{mn}=\frac{\sqrt{p}}{d} \text{ for } (m,n) \in S.
\end{align}

We then calculate the corresponding DP noise operator $F^{\rm DP}_{ab,a'b'} = E_{ab} E_{a'b'}^T$
\begin{equation}
F_{ab,a'b'}^{\rm DP} \!=\! \begin{cases}
a_{00}^2 \sum_{j=0}^{d-1} \ket{j}\bra{j}
& \text{for } a=b=0 \\
& \text{\& } a'=b'=0, \\[8pt]
a_{00} a_{a'b'} \sum_{j=0}^{d-1} \omega_d^{ja'} \ket{j\oplus b'}\bra{j}
& \text{for } a=b=0 \\
& \text{\& } (a',b') \in S, \\[8pt]
a_{00} a_{ab} \sum_{j=0}^{d-1} \omega_d^{ja} \ket{j}\bra{j\oplus b}
& \text{for } (a,b) \in S \\
& \text{\& }  a'=b'=0, \\[8pt]
a_{ab} a_{a'b'} \sum_{j=0}^{d-1} \omega_d^{j(a'+a)+a'(b-b')}
& \text{for } (a,b) \in S \\
\hspace{2cm}\ket{j}\bra{j \oplus (b-b')}
& \text{\& }  (a',b') \in S.
\end{cases}
\end{equation}

Then, after Alice performs the measurement with $\Pi_x^y$, Bob's particle qudit $B$ collapses to (See Appendix \ref{appD3} for more details)
\begin{align}
    &\rho_{B,\rm DP}^{\Pi_x^y} = \sum_{j=0}^{d-1} \frac{1}{d} \ket{j}\bra{j} + \sum_{j_1\neq j_2=1}^{d-1} \sum_{l=0}^{d-1} e^{i\pi \frac{(j_2-j_1) [x(j_1+j_2-d)+2y]}{d}} \nonumber\\
    &\hspace{1cm} \frac{1}{d}(1-p)^2 |\rho_{j_1\ominus l,j_2\ominus l}| \ket{j_1}\bra{j_2},
\end{align}
where $\frac{1}{d}$ is the probability of measurement operator $\Pi_x^y$. The $l_1$-norm coherence reduces to
\begin{align}
    \mathcal{C}_{l_1}(\rho_{B,\rm DP}^{\Pi_x^y}) &= (1-p)^2 \mathcal{C}_{l_1}(\rho_T).
\end{align}
The corresponding efficiency of QC teleportation for DP noise is
\begin{align}
    \eta_{l_1,\rm DP}= (1-p)^2.
\end{align}

\begin{figure}[ht]
\centering
\includegraphics[width=1\linewidth]{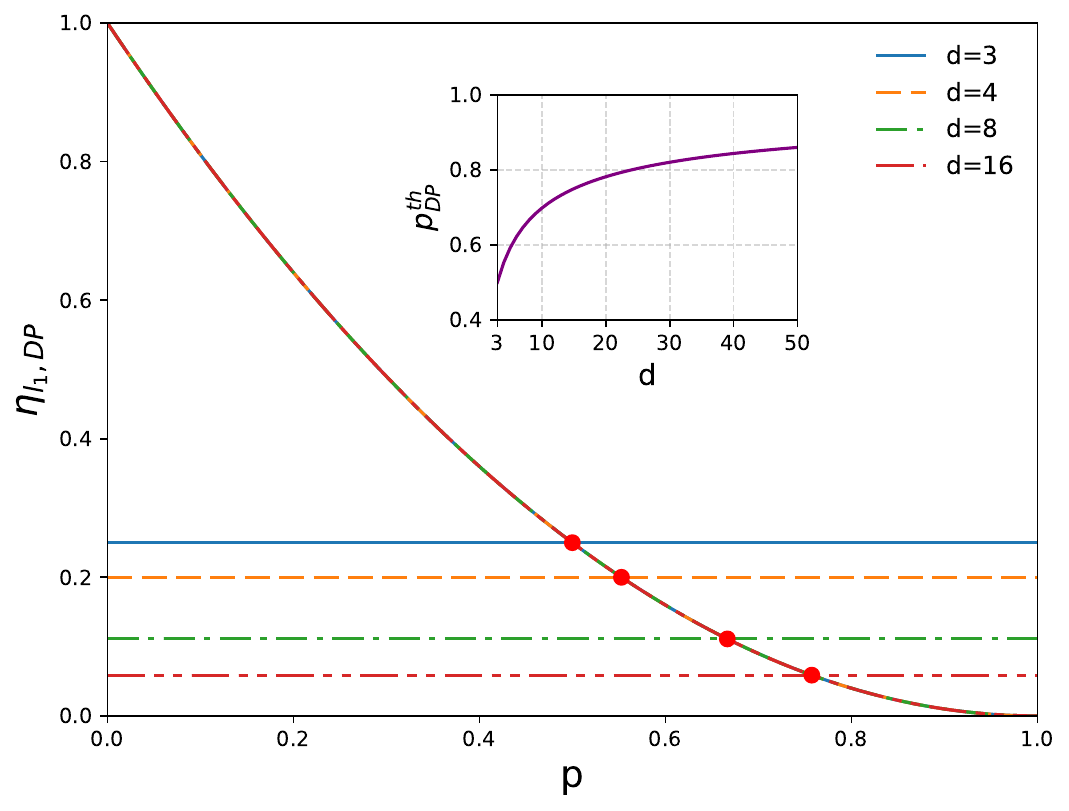}
\caption{The efficiency of QC teleportation for qudits of varying dimensions under DP noise. The horizontal lines corresponds to classical bounds. The red dots indicate the positions of $p_{\rm DP}^{\rm th}$. The inset: The noise threshold $p_{\rm DP}^{\rm th}$ as a function of dimensionality $d$.}
\label{DP}
\end{figure}

The impact of DP noise on the REHDCT protocol is illustrated in Fig. \ref{DP}. A striking observation is that the coherence teleportation efficiency $\eta_{l_1, \rm DP} = (1-p)^2$ is entirely independent of the system dimensionality $d$. This stands in sharp contrast to standard quantum teleportation, where the state fidelity $\langle F\rangle_{\rm DP} = 1 - \frac{d - 1}{d} p$ degrades more severely as $d$ increases \cite{noise2}. The physical mechanism underlying this dimensionality independence lies in the nature of DP noise and the properties of the coherence resource. DP noise can be viewed as an isotropic process that transforms any input state into a completely mixed state with probability $p$. Since the completely mixed state possesses zero coherence under the $l_1$-norm measure, each qudit interaction with the DP channel uniformly scales the original coherence by a factor of $1-p$ without introducing spurious coherence components. In our REHDCT framework, because both particles $A$ and $B$ are subjected to local DP noise, the total scaling factor for the teleported coherence is $(1-p)^2$. Unlike state fidelity, which must account for the increasing difficulty of finding the correct state in a larger Hilbert space, the $l_1$-norm coherence measure focuses on the relative magnitude of off-diagonal elements, which remains invariant to dimensional scaling under DP noise.

Considering that the classical bound decreases with increasing dimensionality, it is noteworthy that high-dimensional quantum systems can still maintain a quantum advantage even in more relaxed DP noise environments.
The noise threshold in this case is given by
\begin{align}
p_{\rm DP}^{\rm th}= 1 - \frac{1}{\sqrt{d + 1}}.
\label{placeholder}
\end{align}
It means that the REHDCT protocol, underpinned by initial phase engineering, allows high-dimensional systems to maintain a superior quantum advantage even under increasingly severe depolarizing environments.

\subsubsection{DF noise}
DF noise refers to perturbations that induce a random jump between $d$ basis states with a probability $p$, which is equivalent to probabilistically applying the displacement operator $X^{m}$, where $X^{m}\ket{j}=\ket{j\oplus m}$. In the symmetric noise model, it is commonly assumed that these $d-1$ types of transitions occur with equal probability, each having a probability of $\frac{p}{d-1}$.
The associated Kraus operators can be expressed as \cite{noise2, noise4}
\begin{align}
    E_{0} = a_0 \sum_{j=0}^{d-1} \ket{j}\bra{j},
    E_{m} = a_m \sum_{j=0}^{d-1} \ket{j}\bra{j \oplus m},
\end{align}
where the coefficients are defined by $a_0=\sqrt{1-p}$, $a_m=\sqrt{p/(d-1)}$ for $m=1, ..., d-1$.

The corresponding operators $F^{\rm DF}_{ab}$ read
\begin{align}
    F_{ab}^{\rm DF} = \left\{
    \begin{aligned}
    & a_0^2 \sum_{j=0}^{d-1} \ket{j}\bra{j} && \text{for } a=0,b=0; \\
    & a_0 a_b \sum_{j=0}^{d-1} \ket{j\oplus b}\bra{j} && \text{for } a=0,b>0; \\
    & a_0 a_a \sum_{j=0}^{d-1} \ket{j}\bra{j\oplus a} && \text{for } a>0,b=0; \\
    & a_a a_b \sum_{j=0}^{d-1} \ket{j\oplus (b-a)}\bra{j} && \text{for } a>0,b>0.
    \end{aligned}
    \right.
\end{align}
Then, after Alice performs the measurement with $\Pi_x^y$, Bob's particle qudit $B$ collapses to (See Appendix \ref{appD4} for more details)
\begin{align}
\rho_{B,\rm DF}^{\Pi_x^y} &= \sum_{j=0}^{d-1} \frac{1}{d} \ket{j}\bra{j} + \sum_{j_1\neq j_2=0}^{d-1} e^{i\pi \frac{(j_2-j_1) [x(j_1+j_2-d)+2y]}{d}} \nonumber\\
&\hspace{1cm}  \frac{Q}{d} \sum_{l=0}^{d-1} \rho_{j_1\ominus l,j_2\ominus l} \ket{j_1}\bra{j_2},
\end{align}
where $1/d$ represents the probability associated with the measurement operator  $\Pi_x^y$. The quantity $Q$ is defined as follows: when $x=0$, $Q=1$; for all other cases $x>0$, $Q=\left(1 - \frac{d}{d-1}p\right)^2$. The coherence is determined through the calculation of the $l_1$-norm, which is given by
\begin{align}
\mathcal{C}_{l_1}(\rho_{B,\rm DF}^{\Pi_x^y}) =
\begin{cases}
\mathcal{C}_{l_1}(\rho_T), & x = 0, \\
\left(1 - \frac{d}{d-1}p\right)^2 \mathcal{C}_{l_1}(\rho_T). & x \neq 0.
\end{cases}
\end{align}
The efficiency of QC teleportation in DF noise is given by
\begin{align}
\eta_{l_1,\rm DF} =
\begin{cases}
1, & x = 0, \\
\left(1 - \frac{d}{d-1}p\right)^2. & x \neq 0.
\end{cases}
\end{align}

The performance of the REHDCT protocol under DF noise reveals a unique and strategically significant feature: the ability to achieve noise-immune coherence teleportation through the optimal selection of the measurement basis. When Alice chooses a POVM set with $x \neq 0$, the results for efficiency $\eta_{l_1, \rm DF}$ and the noise threshold $p_{\rm DF}^{\rm th}$ are analytically equivalent to those observed under PF noise, as given by Eqs. (\ref{eq44}) and (\ref{eq45}), which are illustrated in Fig. \ref{PF}. In these cases, the protocol maintains a quantum advantage that expands with dimensionality $d$, but absolute coherence is still subject to environmental decay.

However, a remarkable result emerges when Alice selects the $x=0$ measurement basis. In this specific configuration, we demonstrate that perfect coherence teleportation ($\eta = 1$) can be maintained regardless of the noise strength $p$. This phenomenon underscores that adapting the measurement basis in accordance with the specific noise symmetry is a powerful strategy for mitigating environmental decoherence. While the target state remains sensitive to noise, the REHDCT protocol---underpinned by initial phase engineering---effectively decouples the coherence resource from the deleterious effects of the DF channel. In the following section, we will focus on identifying the perfect measurement basis that can effectively circumvent the effects of noise.

\subsection{Perfect measurement basis}
\label{sec3C}
To address the aforementioned problem, we must revisit Eq. (\ref{eq22}). By transforming the noise and measurement processes acting on two distinct two-body Hilbert spaces ($\mathcal{H}_{AB}$ and $\mathcal{H}_{TA}$) into a common single-body Hilbert space $\mathcal{B}(\mathcal{H})$, Eq. (\ref{eq22}) presents the entire quantum teleportation process as comprised of two mappings, $\mathcal{G}$ and $ \mathcal{J}_{E}^*$, applied to the target state. Here, $\mathcal{G}$ characterizes the evolution of the shared entangled state, while $ \mathcal{J}_{E}^*$ describes the measurement process. Mathematically, we can conceptualize the effect of noise as acting upon the measurement basis. Therefore, if there exists a set of measurement operators that remains entirely unaffected by noise, we name the corresponding basis as a perfect measurement basis. Obviously, such a condition implies that the following equation holds
\begin{align}
\label{eq60}
    \mathcal{G}_{X}\mathcal{J}_E^* = \mathcal{J}_{E}^*.
\end{align}
The trivial case of Eq. (\ref{eq60}) is that $\mathcal{G}_{X}$ is an identity mapping, which corresponds to the noise-free teleportation of QC. However, there exist certain non-trivial scenarios in which the aforementioned equation holds true. An example of this is DF noise. To elucidate this point, we will re-express the left hand of Eq. (\ref{eq60}) using operators $F_{ab}^{X}$ and $W_{xl\oplus y,l}$ as follows
\begin{align}
    (\mathcal{G}_{X}\mathcal{J}_E^*)_L :=\sum_{l} F_{ab}^{X} W_{xl\oplus y,l}.
\end{align}

In the context of DF noise and the selection of the first set of measurement basis $\Pi_0^y$, the situation simplifies to
\begin{align}
    \sum_{l} F_{00}^{\rm DF} W_{y,l} &= a_0^2 \sum_{l} W_{y,l}, \nonumber \\
    \sum_{l} F_{0b}^{\rm DF} W_{y,l} &= a_0 a_b \sum_{l} W_{y,b\oplus l} = a_0 a_b \sum_{l} W_{y,l}, \\
    \sum_{l} F_{a0}^{\rm DF} W_{y,l} &= a_a a_0 \sum_{l} W_{y,a\ominus l} = a_a a_0 \sum_{l} W_{y,l}, \nonumber\\
    \sum_{l} F_{a'b'}^{\rm DF} W_{y,l} &= a_{a'} a_{b'} \sum_{l} W_{y,b\ominus (a+l)} = a_{a'} a_{b'} \sum_{l} W_{y,l}. \nonumber
\end{align}
Considering the normalization between the coefficients $a_0$ and $a_m$ ($m=a,\ b$), we immediately get
\begin{align}
    \frac{1}{d}\sum_{l,a,b} F_{ab}^{\rm DF} W_{y,l} \rho (F_{ab}^{\rm DF}W_{y,l})^\dagger=\frac{1}{d}\sum_{l}W_{y,l} \rho (W_{y,l})^\dagger.
\end{align}
It is evident that $W_{y,l}=\sum_{j=0}^{d-1} \frac{1}{\sqrt{d}}e^{-i2\pi \frac{jy}{d}}\ket{j\oplus l}\bra{j}$ is independent of the DF noise. Furthermore, DF noise is not the only example; in our previous work \cite{noise1}, we identified that bit-phase flip (BPF) noise also satisfies this condition. Consequently, by selecting an appropriate measurement basis, perfect teleportation of coherence can be achieved.
This conclusion can obviously be extended to the scenario where only a single particle $A$ or $B$ is subjected to DF or BPF noise.
Although we have identified two specific cases, the pursuit of a universal solution to Eq. (\ref{eq60}) may lie beyond the scope of this study.

\section{\label{sec4}Conclusions}

In summary, we have developed a resource-efficient protocol (REHDCT) for the teleportation of quantum coherence in high-dimensional systems. By establishing a theoretical framework comprising $d$ sets of specialized POVM bases, we have demonstrated that utilizing any single POVM set from the framework is sufficient to scale the measurement complexity down from $O(d^2)$ to $O(d)$ while reducing the classical communication cost by 50\% (requiring only $\log_2 d$ bits). Although the REHDCT protocol is mathematically applicable to arbitrary qudit states, the teleported coherence may be degraded for states with generic phase profiles. This degradation originates from the phase-dependent cancellation among density matrix elements during the measurement-induced summation process, which we identify as destructive interference. To mitigate this, we have introduced initial phase engineering as a matching mechanism to ensure that all off-diagonal terms interfere purely constructively, thereby saturating the theoretical upper bound of coherence teleportation for both pure and mixed states. Furthermore, our quantitative analysis of phase deviations confirms the protocol's remarkable robustness; numerical simulations indicate that even with a phase deviation of $\delta_\phi = 0.1$ rad, the average teleportation efficiency exceeds 99.6\% for $d=16$, validating the practical viability of REHDCT for hardware-constrained quantum networks.

We have derived the classical bound on coherence teleportation in the absence of entanglement and established the efficiency of coherence teleportation as a benchmark for comparison. Furthermore, we have evaluated the protocol's performance under various noise channels (AD, PF, DP, and DF) and determined the noise thresholds required to maintain a quantum advantage over the classical benchmark given by $\eta_{l_1, \rm cl} = 1/(d+1)$. The results demonstrate that as the system dimensionality increases, the threshold for preserving quantum advantage approaches the theoretical limit $p \to 1$ under several noise models. Notably, in the case of DF noise, perfect teleportation of QC can be restored simply by selecting the POVM basis with $x = 0$, thereby demonstrating the concept of a ``perfect measurement basis" that is immune to the action of the noise.

The present results underline the practical merit of encoding coherence in high-dimensional Hilbert spaces: the larger the dimension, the lower the classical bound and the higher the resilience against decoherence. This observation aligns with recent experimental progress in photonics \cite{mp1, PRL.123.070505}, where high-dimensional quantum teleportation have already been demonstrated. These findings also provide a practical framework for resource-constrained high-dimensional quantum communication, where minimizing classical bandwidth and measurement complexity is of primary importance.

\begin{acknowledgments}
This work was supported by the Funds of the National Natural Science Foundation of China under Grant Nos. 12365003, 12265004, and Jiangxi Provincial Natural Science Foundation under Grant No. 20242BAB26010. This work was also supported by Jiangxi Provincial Key Laboratory of Multidimensional Intelligent Perception and Control of China (No. 2024SSY03161).
\end{acknowledgments}

\section*{Data Availability}
The data that support the findings of this article are not publicly available. The data are available from the authors upon reasonable request.

\appendix
\section{\label{appA}Derivation of Eq. (\ref{eq8})}
For any arbitrary state of a qudit $\rho_T$, and considering Alice and Bob's shared maximally entangled state $\rho_{AB}=\ket\Phi_{AB} \bra\Phi$,  the resulting collapsed state of particle $B$ following Alice's measurement, which corresponds to the element $\Pi_x^y$ on particles $A$ and $B$ can be expressed as
\begin{align}
    &\rho_B^{\Pi_x^y} = \frac{1}{p^{\Pi_x^y}}{\rm Tr} _{AT}\Big[\Pi_x^y\big(\rho_T\otimes\rho_{AB}]\Big)(\Pi_x^y)^{\dagger}\Big] \nonumber\\
                     &= \frac{1}{p^{\Pi_x^y}}{\rm Tr} _{AT}\Big[\Big(\sum^{d-1}_{l=0} \ket{\Psi_{xl\oplus y,l}} \bra{\Psi_{xl\oplus y,l}}\Big) \Big(\rho_T\otimes \rho_{AB}\Big) \nonumber\\
                     &\hspace{2.2cm}\Big(\sum^{d-1}_{l=0} \ket{\Psi_{xl\oplus y,l}} \bra{\Psi_{xl\oplus y,l}}\Big)^{\dagger}\Big].
\end{align}

Since the basis vectors $\ket{\Psi_{xl\oplus y,l}}$ of the POVM operator $\Pi_x^y$ are orthogonal to each other, the above equation can be written as
\begin{align}
    \rho_B^{\Pi_x^y} = \frac{1}{p^{\Pi_x^y}}\sum^{d-1}_{l=0}\Big[ \sum_{m=0}^{d-1} e^{-i2\pi \frac{(xl\oplus y)m}{d}}\frac{1}{\sqrt{d}} \bra{m}\otimes \bra{m\oplus l}& \nonumber\\
    \Big(\sum_{j,j'=0}^{d-1} \rho_{jj'} \ket{j}\bra{j'}\otimes \sum_{k,k'=0}^{d-1}\frac{1}{d} \ket{kk}\bra{k'k'}\Big)& \nonumber\\
    \sum_{m'=0}^{d-1} e^{i2\pi \frac{(xl\oplus y)m'}{d}} \frac{1}{\sqrt{d}} \ket{m'}\otimes \ket{m'\oplus l} \Big]& \nonumber\\
    = \frac{1}{d^2p^{\Pi_x^y}} \sum_{l,j,j'=0}^{d-1} e^{i2\pi \frac{(xl\oplus y)(j'-j)}{d}} \rho_{jj'} \ket{j\oplus l} \bra{j'\oplus l},
\end{align}
and the corresponding probability of measurement $\Pi_x^y$ is
\begin{align}
    p^{\Pi_x^y} &= {\rm Tr} \Big[\Pi_x^y\Big(\rho_T\otimes\rho_{AB}\Big)(\Pi_x^y)^{\dagger}\Big] \nonumber\\
    &= {\rm Tr} \Big[\frac{1}{d^2} \sum_{l,j,j'=0}^{d-1} e^{i2\pi \frac{(xl\oplus y)(j'-j)}{d}} \rho_{jj'} \ket{j\oplus l} \bra{j'\oplus l}\Big] \nonumber\\
    &= \frac{1}{d^2} \sum^{d-1}_{l=0} \Big(\sum^{d-1}_{j=0}|\rho_{jj}| \Big)
    = \frac{1}{d}.
\end{align}

Finally, the collapsed state held by Bob with respect to the measurement basis \{$\Pi_x^y$\} is given by
\begin{align}
    &\rho_B^{\Pi_x^y} = \frac{1}{d^2p^{\Pi_x^y}} \sum_{l,j,j'=0}^{d-1} e^{i2\pi \frac{(xl\oplus y)(j'-j)}{d}} \rho_{jj'} \ket{j\oplus l} \bra{j'\oplus l} \nonumber\\
        &= \sum_{j=0}^{d-1} \frac{\rho_{jj}}{d} + \sum_{j_1\neq j'_1}^{d-1} \sum_{l=0}^{d-1} e^{i2\pi \frac{(xl\oplus y)(j'_1-j_1)}{d}} \frac{\rho_{j_1j_1'}}{d}  \ket{j_1\oplus l} \bra{j'_1\oplus l}.
\end{align}
For a given set of POVM measurement basis $x$,
the states resulting from different outcomes, labeled by $y = 0, \ldots, d-1$, are obtained by applying the corresponding transformations. Consequently, it can be observed that the only distinctions among the various outcomes reside in the exponential terms.

\section{\label{appB}Numerical Methodology for Random State Sampling}
To ensure a comprehensive and unbiased evaluation of the REHDCT protocol's robustness, we employ Monte Carlo simulations to calculate the average coherence teleportation efficiency $\eta$ over the qudit state space. This approach allows us to quantify the impact of phase deviations $\delta_\phi$ across a diverse ensemble of potential target states.

Pure states are sampled according to the Haar measure on the unitary group $U(d)$, which provides a uniform distribution over the complex projective space. Numerically, a Haar-random pure state $\ket{\psi}$ is generated by drawing a $d$-dimensional complex vector $\mathbf{z} = (z_1, z_2, \dots, z_d)^T$, where each component $z_i$ is an independent and identically distributed (i.i.d.) complex Gaussian random variable with zero mean and unit variance ($z_i \sim \mathcal{CN}(0,1)$). The resulting vector is normalized as:
\begin{align}
\ket{\psi} = \frac{\mathbf{z}}{|\mathbf{z}|} = \frac{1}{\sqrt{\sum_{i=1}^d |z_i|^2}} (z_1, z_2, \dots, z_d)^T.
\end{align}
This procedure ensures that the sampled states are unitarily invariant and cover the Hilbert space uniformly.

Mixed states are sampled according to the Hilbert-Schmidt (HS) measure, which is a standard unitarily invariant measure on the space of density matrices. The HS measure can be physically interpreted as the induced measure obtained by partial tracing a Haar-random pure state of a $d \times d$ composite system. In our numerical implementation, we utilize the Ginibre ensemble to construct these states. A $d \times d$ complex random matrix $G$ is generated with i.i.d. complex Gaussian entries. The corresponding density matrix $\rho$ is then constructed as:
\begin{align}
\rho = \frac{G G^\dagger}{\mathrm{Tr}(G G^\dagger)}.
\end{align}
This construction guarantees that the resulting density matrices are positive semi-definite, normalized ($\mathrm{Tr}(\rho)=1$), and distributed according to the HS measure \cite{mixedstate}.

To achieve reliable statistical results, we adopt an adaptive Monte Carlo sampling strategy. The ensemble size is dynamically increased until the standard error of the mean for the teleportation efficiency $\eta$ converges below a threshold of $10^{-5}$. This level of precision ensures that the results presented in Table \ref{tab1} accurately reflect the protocol's inherent resilience to initial phase deviations rather than numerical fluctuations. All simulations are performed based on the assumption that the protocol's phase-alignment conditions are initially targeted via the engineering process described in Sec.~\ref{sec2B}.

\section{\label{appC}Proof of $\mathcal{E}\otimes \mathcal{I}=\mathcal{I}\otimes \mathcal{E}^T$}
Any CPTP mapping $\mathcal{E}$ can be represented in terms of a set of Kraus operators \cite{QCAQI}. In this context, we denote $A_k$ as the Kraus operator corresponding to the mapping $\mathcal{E}$. Consequently, we have
\begin{align}
    (\mathcal{E}\otimes \mathcal{I})(\ket\Phi_{AB}\bra\Phi)=\sum_k(A_k\otimes I_d)(\ket\Phi_{AB}\bra\Phi)(A_k\otimes I_d)^\dagger.
\end{align}

Considering the following facts
\begin{align}
    &(A_k\otimes I_d)\ket\Phi_{AB} = \sum_i A_k\ket{i} \otimes \ket{i} \nonumber\\
    &= \sum_{i,s,t} (A_k)_{st} \ket s \braket{t|i} \otimes \ket i = \sum_{i,s} (A_k)_{si} \ket{s} \otimes \ket{i} \nonumber\\
    &= \sum_{i,s} \ket{s} \otimes (A_k)_{si}\ket{i} = \sum_{i,s} \ket{i} \otimes(A_k^T)_{si} \ket{s} \nonumber\\
    &= \sum_{i,s,t} \ket{i} \otimes (A_k^T)_{st} \ket{s} \braket{t|i} = \sum_i \ket{i} \otimes A_k^T\ket{i} \nonumber\\
    &= (I_d \otimes A_k^T)\ket\Phi_{AB},
\end{align}
we can obtain the result of Eq. (\ref{eq26})
\begin{align}
\label{eqb3}
    (\mathcal{E}\otimes \mathcal{I})(\ket\Phi_{AB}\bra\Phi) = (\mathcal{I}\otimes \mathcal{E}^T)(\ket\Phi_{AB}\bra\Phi).
\end{align}
Notice that Eq. (\ref{eqb3}) represents that the effect of local operations $\mathcal{E}$ on one part of an entangled system can be mirrored by corresponding operations $\mathcal{E}^T$ on the other part.
Based on this, we can derive the corresponding mapping $\mathcal{G}=\frac{1}{d}\mathcal{E}_B \circ \mathcal{E}_A^T$ for the bipartite entangled state subjected to local noise, and subsequently compute the outcomes under various noisy environments. In the following analysis, we will examine four distinct types of noise environments.

\onecolumngrid

\section{\label{appD}Noisy Teleportation of QC in Qudit}
\subsection{\label{appD1}AD noise}
In the AD noise, after Alice performs  a measurement using the operator $\Pi_x^y$, Bob's state can be determined as follows
\begin{align}
    \mathcal{G}_{\rm AD}\mathcal{J}_E^*(\rho_T) &= \frac{1}{d^2}\sum_{a,b} F_{ab}^{\rm AD} \Big(\sum_{l,j,j'=0}^{d-1} e^{i2\pi \frac{(j'-j)(xl\oplus y)}{d}} \rho_{jj'} \ket{j\oplus l} \bra{j'\oplus l} \Big) (F_{ab}^{\rm AD})^\dagger \nonumber\\
        &= \frac{1}{d^2}\sum_{l=0}^{d-1}\Big( \rho_{d\ominus l,d\ominus l} \ket{0}\bra{0} + \sum_{j=1}^{d-1} (1-p)^2 \rho_{j\ominus l,j\ominus l} \ket{j}\bra{j} + \sum_{j_1'=1}^{d-1} (1-p) e^{i2\pi \frac{j'_1(xl\oplus y)}{d}} \rho_{d\ominus l,j'_1\ominus l} \ket{0}\bra{j'_1} \nonumber\\
        & + \sum_{j_1=1}^{d-1} (1-p) e^{i2\pi \frac{-j_1(xl\oplus y)}{d}} \rho_{j_1\ominus l,d\ominus l} \ket{j_1}\bra{0} + \sum_{j_2\neq j'_2=1}^{d-1} (1-p)^2 e^{i2\pi \frac{(j'_2-j_2)(xl\oplus y)}{d}} \rho_{j_2\ominus l,j'_2\ominus l} \ket{j_2}\bra{j'_2} \nonumber\\
        & + \sum_{b=1}^{d-1} p(1-p) \rho_{d\ominus l,d\ominus l} \ket{b}\bra{b} + \sum_{a=1}^{d-1} p(1-p) \rho_{a\ominus l,a\ominus l} \ket{0}\bra{0} + \sum_{a'=b'=1}^{d-1} p^2 \rho_{d\ominus l,d\ominus l} \ket{0}\bra{0} \Big) \nonumber\\
        &= \frac{1}{d^2}\Big[\Big(1+p(d-1)\Big)\ket{0}\bra{0} + \sum_{j=1}^{d-1} (1-p) \ket{j}\bra{j} + \sum_{l=0}^{d-1} \Big( \sum_{j'_1=1}^{d-1} (1-p) e^{i2\pi\frac{j'_1(xl\oplus y)}{d}} \rho_{d\ominus l,j'_1\ominus l} \ket{0}\bra{j'_1} \nonumber\\
        &+ \sum_{j_1=1}^{d-1} (1-p) e^{i2\pi\frac{-j_1(xl\oplus y)}{d}} \rho_{j_1\ominus l,d\ominus l} \ket{j_1}\bra{0}
        + \sum_{j_2\neq j_2'=1}^{d-1} (1-p)^2 e^{i2\pi\frac{(j'_2-j_2)(xl\oplus y)}{d}} \rho_{j_2\ominus l,j'_2\ominus l} \ket{j_2}\bra{j'_2} \Big)\Big].
\end{align}

And the corresponding probability of success of the measurement is
\begin{align}
    {\rm Tr} [\mathcal{G}_{\rm AD}\mathcal{J}_E^*(\rho_T)] = \frac{1+p(d-1) + (d-1)(1-p)}{d^2} = \frac{1}{d}.
\end{align}

The coherence of the final state $\rho_{B,\rm AD}^{\Pi_x^y}=\frac{1}{d}\mathcal{G}_{\rm AD}\mathcal{J}_E^*(\rho_T)$ is given by
\begin{align}
    \mathcal{C}_{l_1}(\rho_{B,\rm AD}^{\Pi_x^y}) &= \frac{(1-p)}{d} \Big( \sum_{j'_1=1}^{d-1} \Big|\sum_{l=0}^{d-1} (e^{i2\pi\frac{j'_1(xl\oplus y)}{d}} \rho_{d\ominus l,j'_1\ominus l})\Big| + \sum_{j_1=1}^{d-1} \Big|\sum_{l=0}^{d-1} (e^{i2\pi\frac{-j_1(xl\oplus y)}{d}} \rho_{j_1\ominus l,d\ominus l})\Big|\Big) \nonumber\\
    &+ \frac{(1-p)^2}{d} \sum_{j_2\neq j'_2=1}^{d-1}\Big|\sum_{l=0}^{d-1} (e^{i2\pi\frac{(j'_2-j_2)(xl\oplus y)}{d}} \rho_{j_2\ominus l,j'_2\ominus l})\Big| \nonumber\\
    &=\sum_{l=0}^{d-1} \Big( \frac{2(1-p)}{d} \sum_{j_1=1}^{d-1} |\rho_{j_1\ominus l,d\ominus l}| + \frac{(1-p)^2}{d} \sum_{j_2\neq j'_2=1}^{d-1} |\rho_{j_2\ominus l,j'_2\ominus l}| \Big) \nonumber\\
    &= \frac{2(1-p)+(d-2)(1-p)^2}{d} \mathcal{C}_{l_1}(\rho_T).
\end{align}

\subsection{\label{appD2}PF noise}
In the PF noise, after Alice performs  a measurement using the operator $\Pi_x^y$, Bob's state can be determined as follows
\begin{align}
    &\mathcal{G}_{\rm PF}\mathcal{J}_E^*(\rho_T) = \frac{1}{d^2} \sum_{a,b} F_{ab}^{\rm PF} \Big(\sum_{l,j,j'=0}^{d-1} e^{i2\pi \frac{(j'-j)(xl\oplus y)}{d}} \rho_{jj'} \ket{j\oplus l} \bra{j'\oplus l} \Big) (F_{ab}^{\rm PF})^\dagger \nonumber\\
        &= \frac{1}{d^2}\sum_{l,j_1,j_2=0}^{d-1} \Big(a_0^4  + \sum_{b=1}^{d-1} a_0^2 a_b^2 \omega_d^{b(j_1-j_2)} + \sum_{a=1}^{d-1} a_0^2 a_a^2 \omega_d^{a(j_1-j_2)} + \sum_{a',b'=1}^{d-1} a_{a'}^2 a_{b'}^2 \omega_d^{(a'+b')(j_1-j_2)}\Big) e^{i2\pi \frac{(j_2-j_1)(xl\oplus y)}{d}} \rho_{j_1\ominus l,j_2\ominus l} \ket{j_1}\bra{j_2} \nonumber\\
        &= \frac{1}{d^2} \Big[\sum_{j=0}^{d-1} \ket{j}\bra{j} + \sum_{j_1\neq j_2=0}^{d-1} \sum_{l=0}^{d-1} \Big((1-p)^2-\frac{2p(1-p)}{d-1}+\frac{p^2}{(d-1)^2}\Big) e^{i2\pi \frac{(j_2-j_1)(xl\oplus y)}{d}} \rho_{j_1\ominus l,j_2\ominus l} \ket{j_1}\bra{j_2} \Big] \nonumber\\
        &= \frac{1}{d^2}\Big[\sum_{j=0}^{d-1} \ket{j}\bra{j} + \sum_{j_1\neq j_2=0}^{d-1} \sum_{l=0}^{d-1} (1-\frac{d}{d-1}p)^2 e^{i2\pi \frac{(j_2-j_1)(xl\oplus y)}{d}} \rho_{j_1\ominus l,j_2\ominus l} \ket{j_1}\bra{j_2} \Big].
\end{align}

And the corresponding probability of success of the measurement is
\begin{align}
      {\rm Tr} [\mathcal{G}_{\rm PF}\mathcal{J}_E^*(\rho_T)] =\frac{1}{d}.
\end{align}

The coherence of the final state $\rho_{B,\rm PF}^{\Pi_x^y}=\frac{1}{d}\mathcal{G}_{\rm PF}\mathcal{J}_E^*(\rho_T)$ is given by
\begin{align}
    \mathcal{C}_{l_1}(\rho_{B,\rm PF}^{\Pi_x^y}) &= \frac{1}{d}\sum_{j_1\neq j_2=0}^{d-1} \Big|\sum_{l=0}^{d-1} (1-\frac{d}{d-1}p)^2 e^{i2\pi \frac{(j_2-j_1)(xl\oplus y)}{d}} \rho_{j_1\ominus l,j_2\ominus l}\Big| \nonumber\\
    &= (1-\frac{d}{d-1}p)^2 \mathcal{C}_{l_1}(\rho_T).
\end{align}

\subsection{\label{appD3}DP noise}
In the DP noise, after Alice performs  a measurement using the operator $\Pi_x^y$, Bob's state can be determined as follows
\begin{align}
    &\hspace{0.5cm}\mathcal{G}_{\rm DP}\mathcal{J}_E^*(\rho_T) \nonumber\\
    &= \frac{1}{d^2} \sum_{a,b,a',b'} F_{ab,a'b'}^{\rm DP} \Big(\sum_{l,j,j'=0}^{d-1} e^{i2\pi \frac{(j'-j)(xl\oplus y)}{d}} \rho_{jj'} \ket{j\oplus l} \bra{j'\oplus l} \Big) (F_{ab,a'b'}^{\rm DP})^\dagger \nonumber\\
    &= \frac{1}{d^2} \sum_{l,j_1,j_2=0}^{d-1} \Big(a_{00}^4 \rho_{j_1\ominus l,j_2\ominus l} + \sum_{(a,b)\in S}^{d-1} a_{00}^2 a_{ab}^2 \omega_d^{a(j_1-j_2)} \rho_{j_1\oplus (b-l),j_2\oplus (b-l)} + \sum_{(a',b')\in S}^{d-1} a_{00}^2 a_{a'b'}^2 \omega_d^{a'(j_1-j_2)} \rho_{j_1\ominus(b'+l),j_2\ominus(b'+l)} \nonumber\\
    &+ \sum_{(a_1,b_1),(a'_1,b'_1)\in S}^{d-1} a_{a_1 b_1}^2 a_{a_1'b_1'}^2 \omega_d^{(a_1+a'_1)(j_1-j_2)} \rho_{j_1\oplus (b_1-b'_1-l),j_2\oplus (b_1-b'_1-l)} \Big)e^{i2\pi \frac{(j_2-j_1)(xl\oplus y)}{d}} \ket{j_1}\bra{j_2}\nonumber\\
    &= \frac{1}{d^2} \sum_{j=0}^{d-1} \ket{j}\bra{j} + \sum_{j_1\neq j_2=1}^{d-1} \sum_{l=0}^{d-1} e^{i\pi \frac{(j_2-j_1) [x(j_1+j_2-d)+2y]}{d}} \Big((1-\frac{d^2-1}{d^2}p)^2
    + \sum_{(a,b)\in S}^{d-1} e^{i2\pi \frac{(a-bx)(j_1-j_2)}{d}} (1-\frac{d^2-1}{d^2}p) \frac{p}{d^2}   \nonumber\\
    &+  \sum_{(a',b')\in S}^{d-1} e^{i2\pi \frac{(a'+b'x)(j_1-j_2)}{d}} (1-\frac{d^2-1}{d^2}p) \frac{p}{d^2} + \sum_{(a_1,b_1),(a'_1,b'_1)\in S}^{d-1} e^{i2\pi \frac{[(a'_1-a_1)+x(b'_1-b_1)](j_1-j_2)}{d}} \frac{p^2}{d^4} \Big)|\rho_{j_1\ominus l,j_2\ominus l}| \ket{j_1}\bra{j_2} \nonumber\\
    &= \frac{1}{d^2}\Big[\sum_{j=0}^{d-1} \ket{j}\bra{j} + \sum_{j_1\neq j_2=1}^{d-1} \sum_{l=0}^{d-1} e^{i\pi \frac{(j_2-j_1) [x(j_1+j_2-d)+2y]}{d}} (1-p)^2 |\rho_{j_1\ominus l,j_2\ominus l}| \ket{j_1}\bra{j_2}\Big].
\end{align}

And the corresponding probability of success of the measurement is
\begin{align}
      {\rm Tr} [\mathcal{G}_{\rm DP}\mathcal{J}_E^*(\rho_T)] =\frac{1}{d}.
\end{align}

The coherence of the final state $\rho_{B,\rm DP}^{\Pi_x^y}=\frac{1}{d}\mathcal{G}_{\rm DP}\mathcal{J}_E^*(\rho_T)$ is given by
\begin{align}
    \mathcal{C}_{l_1}(\rho_{B,\rm DP}^{\Pi_x^y}) = (1-p)^2 \mathcal{C}_{l_1}(\rho_T).
\end{align}

\subsection{\label{appD4}DF noise}
In the DF noise, after Alice performs  a measurement using the operator $\Pi_x^y$, Bob's state can be determined as follows
\begin{align}
    \mathcal{G}_{\rm DF}\mathcal{J}_E^*(\rho_T) &= \frac{1}{d^2} \sum_{a,b} F_{ab}^{\rm DF} \Big(\sum_{l,j,j'=0}^{d-1} e^{i2\pi \frac{(j'-j)(xl\oplus y)}{d}} \rho_{jj'} \ket{j\oplus l} \bra{j'\oplus l} \Big) (F_{ab}^{\rm DF})^\dagger \nonumber\\
        &= \frac{1}{d^2}\sum_{l,j_1,j_2=0}^{d-1} \Big(a_0^4  + \sum_{b=1}^{d-1} a_0^2 a_b^2 \omega_d^{xb(j_2-j_1)} + \sum_{a=1}^{d-1} a_0^2 a_a^2 \omega_d^{-xa(j_2-j_1)} + \sum_{a',b'=1}^{d-1} a_{a'}^2 a_{b'}^2 \omega_d^{x(b'-a')(j_2-j_1)}\Big) \nonumber\\
        &\hspace{2cm}e^{i\pi \frac{(j_2-j_1) [x(j_1+j_2-d)+2y]}{d}} |\rho_{j_1\ominus l,j_2\ominus l}| \ket{j_1}\bra{j_2}.
\end{align}

For $x=0$:
\begin{align}
    \mathcal{G}_{\rm DF}\mathcal{J}_E^*(\rho_T) &= \frac{1}{d^2}\sum_{l,j_1,j_2=0}^{d-1} e^{i\pi \frac{(j_2-j_1) [x(j_1+j_2-d)+2y]}{d}} |\rho_{j_1\ominus l,j_2\ominus l}| \ket{j_1}\bra{j_2} \nonumber\\
        &= \frac{1}{d^2}\Big[\sum_{j=0}^{d-1} \ket{j}\bra{j} + \sum_{j_1\neq j_2=0}^{d-1} \sum_{l=0}^{d-1} e^{i\pi \frac{(j_2-j_1) [x(j_1+j_2-d)+2y]}{d}} |\rho_{j_1\ominus l,j_2\ominus l}| \ket{j_1}\bra{j_2}\Big].
\end{align}

For $x\neq 0$:
\begin{align}
    \mathcal{G}_{\rm DF}\mathcal{J}_E^*(\rho_T) &= \frac{1}{d^2}\sum_{l,j_1,j_2=0}^{d-1} (1-\frac{d}{d-1}p)^2 e^{i\pi \frac{(j_2-j_1) [x(j_1+j_2-d)+2y]}{d}} |\rho_{j_1\ominus l,j_2\ominus l}| \ket{j_1}\bra{j_2} \nonumber\\
        &= \frac{1}{d^2}\Big[\sum_{j=0}^{d-1} \ket{j}\bra{j} + \sum_{j_1\neq j_2=0}^{d-1} \sum_{l=0}^{d-1} \left(1 - \frac{d}{d-1}p\right)^2 e^{i\pi \frac{(j_2-j_1) [x(j_1+j_2-d)+2y]}{d}} |\rho_{j_1\ominus l,j_2\ominus l}| \ket{j_1}\bra{j_2}\Big].
\end{align}

The success probability in all cases is given by
\begin{align}
      {\rm Tr} [\mathcal{G}_{\rm DF}\mathcal{J}_E^*(\rho_T)] =\frac{1}{d}.
\end{align}

The coherence of the final state $\rho_{B,\rm DF}^{\Pi_x^y}=\frac{1}{d}\mathcal{G}_{\rm DF}\mathcal{J}_E^*(\rho_T)$ can be expressed as
\begin{align}
\mathcal{C}_{l_1}(\rho_{B,\rm DF}^{\Pi_x^y}) =
\begin{cases}
\mathcal{C}_{l_1}(\rho_T) &  x = 0, \\
\left(1 - \frac{d}{d-1}p\right)^2 \mathcal{C}_{l_1}(\rho_T) & x \neq 0.
\end{cases}
\end{align}

\twocolumngrid
%
\end{document}